\documentclass[fleqn,usenatbib]{mnras}

\usepackage{newtxtext,newtxmath}

\usepackage[T1]{fontenc}
\usepackage{ae,aecompl}

\usepackage{graphicx}	
\usepackage{amsmath}	

\title[Accretion rates onto AMSs]{Numerical modeling the mass feeding rates onto accretion-modified stars embedded within AGN disks}

\author[Luo \& Wang]{
Yang~Luo,$^{1}$\thanks{E-mail: luoyang@ynu.edu.cn (YL)}
Jian-Min~Wang,$^{2,3,4}$
\\
$^{1}$Department of Astronomy, and Key Laboratory of Astroparticle Physics of Yunnan Province, Yunnan University, Kunming, Yunnan 650091, China\\
$^{2}$Key Laboratory for Particle Astrophysics, Institute of High Energy Physics,
Chinese Academy of Sciences, 19B Yuquan Road, Beijing 100049, China\\
$^{3}$School of Astronomy and Space Sciences, University of Chinese Academy of Sciences, 
19A Yuquan road, Beijing 100049, China\\
$^{4}$National Astronomical Observatory of China, 20A Datun Road, Beijing 100020, China
}
\date{Accepted XXX. Received YYY; in original form ZZZ}

\pubyear{2025}

\begin{document}
\maketitle

\label{firstpage}

\begin{abstract}

Accretion disks surrounding supermassive black holes can potentially form stars within the self-gravitating region. These stars undergo high accretion rates because of the dense environment of the active galactic nuclei (AGN) accretion disk. The vorticity of the AGN disk may influence the ultimate mass feeding rate toward the star. In our study, we simulate mass feeding rates onto stars at different AGN disk thicknesses through $3$D numerical models to explore the relationship between feeding rates and the thermal mass of the star ($q_{\rm th}$), defined as the ratio of the star's Bondi radius to the AGN disk thickness. Our findings indicate that disk shearing with angular momentum can notably decrease the feeding rate, and we provide an approximate formula that links the feeding rate based on the angular momentum of the surrounding gas and the thermal mass $q_{\rm th}$. Lastly, we examine the potential feedback of the rapidly accreting stars on the AGN disk and their subsequent evolution.
 
\end{abstract}

\begin{keywords}
methods: numerical --- quasars: supermassive black holes --- accretion, accretion discs --- stars: black holes --- stars: protostars 
\end{keywords}


\section{Introduction}

The accretion power released via an accretion disk surrounding a central supermassive black hole (SMBH) is widely believed to be responsible for the observations of active galactic nuclei (AGN) \citep{pringle81,rees84,kormendy13}. The accretion process can be theoretically represented by various models of accretion disks, which are determined by the rate of accretion \citep{shakura73,abramowicz88,narayan94,yuan14,padovani17}. As the accretion disk moves mass towards the AGN center, it tends to show self-gravitational effects at its outer edges \citep{paczynski78,shlosman87,goodman03,lodato07}.

When the density within the AGN disk exceeds the mean density associated with the external gravitational field by a small factor, the local Jeans instability may be triggered. With a further increase in the mass feeding rate, the disk can potentially become a globally self-gravitating disk \citep{shlosman87}. The criterion for identifying this axis-symmetric instability can be expressed through the Toomre parameter $Q$ as $Q<1$ \citep{toomre64}. As the instability arises, the disk will form clumps on a characteristic scale corresponding to the wavelength that grows most rapidly, typically around the local Jeans length. These self-gravitating clumps may contract further and eventually form stars \citep{shlosman89b, goodman04, cantiello21}. 

When the gravitational instability sets in, inducing turbulent heating that counterbalances radiative cooling \citep{paczynski78}, fragments may still form if the disk cooling time is comparable or shorter than the local dynamical timescale \citep{gammie01}. As the fraction of support by radiation pressure increases, the disk becomes more prone to fragmentation \citep{chen23}.

Stars within AGN disks may have originated from the clump collapse caused by disk instability or from the capture of nuclear star clusters \citep{artymowicz93,collin08,wang10,mapelli12,wang21,wang24}. Regardless of their origin, as stars evolve, compact objects such as black holes could form within the AGN accretion disk \citep{dittmann20,ali-dib23}. These stars or compact objects in AGN disks accrete gas from their surroundings and have the potential to grow into supermassive objects \citep{goodman04}. In such a dense environment, this new type of fast accretion populations could be termed as accretion-modified stars (AMSs) \citep{wang21,wang21b}. 

Obtaining direct observational evidence of stars within AGN disks continues to be difficult due to the dominant brightness of the central AGN \citep[e.g.,][]{zhang24,zhou24,xing25}. Nevertheless, observations of stellar clusters near the galactic center \citep[e.g.,][]{nayakshin05,paumard06,bartko10,zhu18,nayakshin18,neumayer20,schodel20,jia23} offer some indirect evidence of stellar capture. Subsequently, stars and compact objects within these clusters could slowly sink towards the SMBHs due to dynamical friction with surrounding gas\citep{gerhard01,collin08, kennedy16}. Despite these difficulties, there is ongoing interest in the theoretical exploration of the growth mechanisms of AMSs within AGN disks. Specifically, what are the potential growth paths for AMSs and what is the expected rate of mass increase?

A straightforward estimation of the mass feeding rate $\dot{M}$ toward an AMS suggests that accretion could exceed the AMS Eddington limit $\dot{M}_{\rm edd}$, with the Eddington ratio potentially reaching $m \equiv \dot{M}/\dot{M}_{\rm edd} \sim 10^9$ \citep{wang21}.  Here, $\dot{M}_{\rm edd}$ is defined as $\dot{M}_{\rm edd} \equiv L_{\rm edd}/c^2$, where $L_{\rm edd}$ represents the Eddington luminosity and $c$ denotes the speed of light. If the effect of angular momentum is ignored, the mass feeding rate is close to the Bondi rate, but with angular momentum, then the rate could be reduced and need detailed investigations. 

The issue of AMS accretion within an AGN disk parallels that of planet accretion within a proto-stellar disk \citep{piso14,lee15,berardo17,ginzburg19}. Early numerical simulations suggest that during the initial phases of planet accretion, the rate is linked to a Bondi-like accretion rate, which is adjusted by a thermal mass parameter denoted as $q_{\rm th}$ \citep{ginzburg19,choksi23}. This parameter $q_{\rm th}$ is defined as the ratio of the planet's Bondi radius to the thickness of the proto-stellar disk. Gas accretion occurs in a runaway phase for $q_{\rm th} < 0.3$, with the maximum accretion rate following a Bondi scaling law expressed as $\dot{M} \propto \rho M_{\rm p}^2/h^3$, where $\rho$ represents the ambient disc density, $M_{\rm p}$ is the planet mass, and $h$ is the aspect ratio of the proto-stellar disk \cite{choksi23}. 

Considering the potential scalability of our AMS accretion models with planet formation, it raises the question of whether the approximations for accretion rates in planet formation remain applicable in this context, especially concerning how the rate is affected by the angular momentum present in the AGN disk. In this work, by performing numerical simulations, we examine the effect of angular momentum on the mass feeding rate onto the central AMS.

\subsection{Theoretical expectations of the AMSs growth}
\label{sect:expect}

Given the rapid mass supply to the AMSs within the AGN disks, how would the AMSs grow in mass? 

If the AMSs originate from the capture of a nuclear star cluster, the gas drag between the star and the gaseous disk would lead to the gradual trapping of stars \citep{artymowicz93, gerhard01, collin08, kennedy16}. Once situated within the disk, the AMSs accumulate gas and grow in mass. 

In the scenario where the AMSs emerge from the gravitational instability of the AGN disks, which undergo runaway gravitational fragmentation, fragments form, and AMSs are expected to follow the evolution of these fragments.

The initial formation and growth of clumps or fragments could be described as the gravitational collapse of a self-gravitating gas sphere. Although magnetic fields and rotation likely play important roles, we can still estimate the mass feeding rate, before the angular momentum becomes important. Initially, the clumps contract under isothermal or quasi-adiabatic conditions, with an increasing imbalance of gravitational forces over pressure forces, leading to an eventual free-fall collapse. Early research suggests that prior to the formation of a core/AMS, an envelope density distribution follows a $r^{-2}$ law (isothermal) or a slightly less steep power-law (quasi-adiabatic) \citep{larson69, penston69}. Before the gas reaches the centrifugal barrier, it flows inwards at a free-fall velocity. Beyond the disk and the centrifugal barrier, the gravitational potential is primarily determined by the clump total mass, allowing us to estimate the mass feeding rate as

\begin{equation}
        \dot{M}_{\rm ff} \sim M_{\rm clump}/t_{\rm ff}, 
\end{equation}

where $M_{\rm clump}$ is expect to the the clump total mass, and the $t_{\rm ff}$ is the gas free-fall timescale. The clumps are formed due to gravitational instability, we can assume the $M_{\rm clump}$ as the local Jeans mass, $M_{\rm clump} \sim (\pi c_{\rm s}^2/G \rho)^{3/2}\pi\rho/6$, where $c_{\rm s}$ is the gas sound speed, $G$ is the gravitational constant, and $\rho$ is the gas density. The free-fall timescale is therefore $t_{\rm ff} \sim (3\pi/32G\rho)^{1/2}$. Finally we obtain the mass feeding rate as 

\begin{equation}
    \dot{M}_{\rm ff} \sim \alpha_{\rm ff} c_{\rm s}^3/G,
\end{equation}

where $\alpha_{\rm ff}$ is the coefficient on the order of one.

As the central core or AMS form and increase in mass, the gravitational potential becomes dominated by the core/AMS mass $M_{\rm ams}$. The gravitational force stemming from the central mass accumulation then starts to impact the gas accretion process. In scenarios where gas falls spherical symmetry onto a central point object, the feeding rate can be approximately formulated using the Bondi rate \citep{bondi52,ruffert94} as 

\begin{align}
\label{equ:bondi}
        \dot{M}_{\rm B} &=  \pi G^2 M_{\rm ams}^2 \frac{\rho}{c_{\rm s}^3}\left(\frac{2}{5-3\gamma}\right)^{(5-3\gamma)/(2\gamma -2)} \\
                           &= \alpha_{\rm B} G^2 M_{\rm ams}^2 \frac{\rho}{c_{\rm s}^3},
\end{align}

where $\rho$ and $c_{\rm s}$ are the gas density and sound speed at the Bondi radius, respectively. We re-write the equation with a coefficient $\alpha_{\rm B}$. Here $\gamma$ is the adiabatic index, and the dependence of the accretion rate on $\gamma$ is rather week. The factor $(2/(5-3\gamma))^{(5-3\gamma)/(2\gamma -2)}$ varies from $2.5$ in the limit $\gamma=1.4$ to $4.5$ in the limit $\gamma=1.0$. 

When taking into account the relative motion between the AMSs and the AGN disk, then the feeding rate could be described as the Hoyle–Lyttleton–Bondi formulation \citep{hoyle39}. Nevertheless, factor such as hydrodynamical drag would reduce the relative motion, causing the relative velocity to drop below the sound speed within the AGN disk \citep{wang21}.

The threshold for the Bondi rate to become the dominant mass feeding rate could be estimated by comparing the two rates $\dot{M}_{\rm ff}$ and $\dot{M}_{\rm B}$. We could find that there is a critical mass, 
\begin{equation}
        M_{\rm crit}^2 \sim \frac{\alpha_{\rm ff}}{\alpha_{\rm B}} \frac{c_{\rm s}^6}{\rho G^3},
\end{equation}
above which mass, the mass feeding rate would be dominated the the central mass potential, and the rate $\dot{M}_{\rm B}$ overwhelm $\dot{M}_{\rm ff}$. The timescale for the free-fall collapse to reach the stage of AMS potential domination is $\tau \sim c_{\rm s}^3/(\alpha_{\rm B} \rho G^2)$.

The growth of AMSs does not always follow the Bondi accretion model, as the gas pressure and rotation could modify the accretion rate. It has been demonstrated that the presence of vorticity near the central mass accumulation can significantly alter both the accretion rate and the gas flow morphology \citep{proga03,krumholz05}. Within the AGN disk, the shearing of the disk imparts angular momentum to the accreting gas \citep{dittmann21}.

In the early work \citep[e.g.,][]{dittmann21}, the mass feeding rate with the impact of angular momentum toward the AMS is estimated as 
\begin{equation}
        \dot{M} = \dot{M}_{\rm B}  f(q_{\rm th}),
\end{equation}

where $f(q_{\rm th})$ represents the suppression factor proposed by \citet{krumholz05}. This factor is derived under the assumption of a uniform background flow with constant specific angular momentum. What might be the form of $f(q_{\rm th})$ in an AGN disk's shearing environment?

When the AMS first forms within the AGN disk or becomes captured and confined within it, the effective accretion radius is significantly smaller than both the thickness of the AGN disk, denoted as $H_{\rm ams}$, and the Hill radius $R_{\rm hill} = (\mu /3)^{1/3} R_{\rm ams}$, where $\mu \equiv M_{\rm ams}/M_{\rm agn}$ represents the ratio of the AMS mass $M_{\rm ams}$ to the AGN SMBH mass $M_{\rm agn}$. Given an opportunity to feed the gas from its vicinity, the AMS may accrete a significant amount of gas, leading to a runaway accretion process due to the mass-dependent nature of the rate. When the Bondi radius of the AMS reaches the AGN disk scale height or the Roche radius, the accretion rate of the AMS deviates from the suppressed Bondi rate. The geometric effects of the AGN disk, and the tidal effects could play a role. When AMS Bondi radius reaches the Roche radius \citep[e.g.,][]{artymowicz93}, the critical mass ratio, as a function of the disk thickness aspect ratio $h\equiv H_{\rm ams}/R_{\rm ams}$, is approximated as

\begin{equation}
        \mu_{\rm crit} = 3^{-1/2}(H_{\rm ams}/R_{\rm ams})^3 = 3^{-1/2}h^3.
\end{equation}

When the AMS reaches the critical mass ratio $\mu_{\rm crit}$, its accretion rate surpasses the Eddington rate $\dot{M}_{\rm Edd} \equiv L_{\rm Edd}/c^2$, where $L_{\rm Edd}$ is the AMS Eddington luminosity. Various processes such as feedback from gas accretion, nuclear radiation of the AMS, or mechanical outflow could influence the growth of the AMS. The exact impact of these processes on the growth of AMS is still under investigation \citep[e.g,][]{cantiello21,dittmann21}.

Considering scalability, and based on the accretion rate results in terms of $\rho R_{\rm ams}^3 \Omega$ from planet accretion and growth within a proto-stellar disk, it is anticipated that the AMS accretion will follow to a Bondi scaling when $\mu/\mu_{\rm crit} \leq 1$ or $q_{\rm th} \leq 0.3$. In this context, the accretion rate is expected to vary as $\dot{M} \sim q_{\rm th}^2 h^3$, here $h = r_{\rm B}/H_{\rm ams}$. Conversely, for $q_{\rm th} \geq 1$, the accretion rate during gap opening will be proportionate to $\sim q_{\rm th} h^3$ \citep{choksi23}.

With the AMS growing in mass, the tidal effects and the angular momentum exchange between the AMS and the gas disk becomes important. In the case of large AMS masses, the angular momentum flux of the AMS locally dominates the viscous flux \citep{kley12}. As a consequence, gas is repelled from high-$m$ resonances, and obtain a net torque due to Lindblad resonances \citep{goldreich80,kley12}. The surface density of the AGN disk drops near $R_{\rm agn}$, forming a gap - an annular region in which the surface density is smaller than its unperturbed value \citep[for an review of related topics, see]{armitage07,kley12}. Together with the geometric effect, the gas supply to the AMS would be significantly slowed down. If the accretion rate drops to a certain level, the potential AMS wind could overcome the inflowing momentum and finally clean up the mass in the gap and cut off the gas supplying from the disk reservoir. 

The size of the gap is determined by the interplay between tidal torques, which remove gas from the gap region, and the viscosity, which can fill the gap. Following the work of \citet{takeuchi96}, we observe that the timescale for viscous diffusion to close a gap of width $\Delta R$ is approximately, $t_{\rm close} \sim \Delta R ^2/\nu$, where $\nu$ is the viscosity coefficient of the AGN disk. On the other hand,  the timescale for opening a gap due to tidal torque at a $m$-th order Lindblad resonance is given by $t_{\rm open} \sim 1/(m^2 \mu^2 \Omega(R_{\rm ams})) (\Delta R/R_{\rm ams})^2$ \citep[e.g.,][]{armitage07}.

By considering typical parameters of an AGN disk, we can make an rough estimate for the both time scales. It appears that AMS can open a gap over a significantly long timescale (see calculations in the Discussion Section \ref{subsect:tidal}). This extended timescale is crucial for the subsequent evolution of the AMS. The AMS could reaches the Hill mass within a few dynamical timescales, which is notably shorter than the lifetime of an AMS. This suggests that the final product of the AMS, such as supernovae or stellar-mass black holes, may form during the gap-opening phase. However, a more detailed investigation of the AMS growth after reaching $\mu_{\rm crit}$ will be the focus of our future research.

In this study, we begin by examining the influence of disk angular momentum on the mass feeding rate onto an AMS, and we introduce an analytical expression for the suppression factor $f(q_{\rm th})$, and the findings are compared with previous planet accretion models. Section \ref{sect:numerical} outlines the numerical techniques employed in our simulations. Results are presented in Section \ref{sect:result}. Lastly, Section \ref{sect:discuss} and \ref{sect:conclusion} are dedicated to discussions and conclusions.

\section{Numerical Method}
\label{sect:numerical}

To model accretion of point particles in a shear flow, a series of 3D simulations was carried out using the Eulerian adaptive mesh refinement (AMR) code Enzo-2.5 \citep{bryan97,norman99,bryan14}\footnote{https://github.com/enzo-project/enzo-dev/tree/gold-standard-v15}. The Eulerian equations of hydrodynamics including gravity, are given in \citet{bryan14}. The hydrodynamical equations are solved employing the Runge-Kutta second-order based on the MUSCL method in combination with an HLL Riemann solver. The code is capable of solving the Poisson equations for gas self-gravity by employing a fast Fourier technique \citep{hockney88} on the root grid and a multigrid technique on subgrids for each time step.  The gravitational acceleration in our models are mainly dominated by the central point source, and the central particle is distributed onto the grids using the second-order cloud-in-cell (CIC) interpolation technique \citep{hockney88} to form a spatially-discretized density field.

The central AMS is represented by a sink particle, positioned at the highest refined level within the domain. This particle experiences the gravitational force similar to those acting on a grid cell at that level of refinement, and is capable of moving through the gas, while accreting material from it \citep{krumholz04}. The positions and velocities of the particle are updated for each time step using a drift-kick-drift algorithm \citep{hockney88}.

The numerical simulations use a local shearing box approach using a Cartesian grid where $x, y, z$ correspond to the radial, azimuthal, and vertical directions in a global disk \citep{goldreich65,hawley95,stone10}. The computational domain is a rectangular prism with sides of $(10\times20\times10) r_{B}$, and a root grid dimension of $64\times128\times64$. The boundary condition for the $x$ direction is set as shearing, while the other two directions are set as periodic. In the rotating shearing box frame, the primary control parameter is the rotation rate $\Omega$, with a shearing factor of $\rm q \equiv -dln\Omega/dlnr = 3/2$. 

The implementation details of the shearing box in Enzo are documented in \citet{zhao2010}. This local shearing box method uses a reference frame at a radius $R_0$, rotating with a velocity of $v_{\rm c} = R_0 \Omega$ \citep{hawley96}. The coordinate system is converted to Cartesian, where the differential rotation is presented as a uniform shear flow described by $v_{\rm y} = -q \Omega x$. With this coordinate transform, the momentum equation becomes

\begin{equation}
    \frac{\partial \rho {\bf v}}{\partial t} + {\bf\nabla\cdot} [ \rho{\bf vv} + p{\sf I} ]  = 
\rho \Omega^{2}(2qx{\bf \hat{i}} - z{\bf \hat{k}}) - 2\Omega {\bf \hat{k}} \times \rho {\bf v},
\label{eq:cons_momentum} \\
\end{equation}
where $p$ is the gas pressure.

Within the shearing box, the momentum equation includes three extra source terms, which account for the radial and vertical gravitational forces in the rotating frame (the first two source terms on the RHS of the equation), as well as the Coriolis force (the third term) \citep{stone10}.

The grid cells are adaptively refined based on the baryon mass in the cells. The grid is refined by a factor of $2$ in length scale if the gas density exceeds $\rho_0 2^{3+\alpha_{\rm ref}}l$, where $\rho_0$ is the mean density of the top grid. Here $l$ is the refinement level and $\alpha_{\rm ref}$ is set to $-0.5$, which makes the refinement super-Lagrangian \citep{bryan14}. In all simulations, we set the maximum refinement level to $7$, which provides approximately $1.2\times 10^{-3} r_{\rm B}$ physical resolution when the maximum refinement level is reached.

We aim for a smooth transition in the accretion rate as the sink particle crosses the cell boundaries \citep{krumholz04}. After conducting experiments of different methods, we have decided to use an accretion rate formula of $\dot{M}_{\rm sink} = 4\pi r_{\rm acc}^2\rho_{\rm a} v_{\rm r,a}$, given as the mass flux toward the sink. Here, the accretion radius $r_{\rm acc}$ is defined as four times the minimum cell size. The parameters $\rho_{\rm a}$ and $v_{\rm r,a}$ represent the averaged density and radial velocity at the accretion radius, respectively. Alternatively, we also explored another approach of accretion rate calculation, utilizing formula $\dot{M}_{\rm sink} = 0.3\pi c_{\rm s,a}^2 \Sigma_{\rm a}/\Omega_{\rm a}$, where $c_{\rm s,a}$, $\Sigma_{\rm a}$, and $\Omega_{\rm a}$ denote the sound speed, surface density and angular velocity at the accretion radius \citep{debuhr10}. Both approaches towards determining the sink rate eventually converge during later stages of evolution, yielding consistent outcomes.

The accretion onto the sink is calculated at each timestep. Subsequently, a corresponding quantity of mass from the accretion region is extracted from the grid and transferred to the sink particle \citep{krumholz04}.

\subsection{initial conditions}

Our approach involves simulating the mass feeding rate onto an AMS within an AGN disk through numerical simulations employing a sink particle positioned at the center of the domain to represent an AMS. This sink particle accretes gas from the surrounding medium, which we model as a gaseous disk. The disk is characterized by a disk with a Toomre parameter $Q$ and a thickness ratio $h$. The gas temperature or sound speed can be determined as $c_{\rm s} = H_{\rm ams}\Omega$, where $\Omega$ represents the angular velocity of the gas shearing. The sound speed is constant with respect to the vertical coordinate $z$. 

The sink particle is set as static, but the surrounding medium moves with a tangential velocity in the azimuthal $y$ direction as $R_{\rm ams} \Omega$. We initialize the AMS with a mass of $M_{\rm ams} = \mu M_{\rm agn}$ and $M_{\rm agn} = 10^8\, \rm M_{\odot}$. In this work, we only consider the early phase of AMS accretion, so the mass ratio $\mu$ is smaller than $\mu_{\rm crit}$, such that the thermal mass parameter $q_{\rm th} \equiv r_{\rm B}/H_{\rm ams}$, is also smaller than one. While for $\mu/\mu_{\rm crit} \sim 1$, the AGN disk's geometry plays a role in influencing the accretion rate towards the sink particle. So in this study, the density profile around the sink particle is assumed that varies with height along the $z$ direction following an exponential disk profile given by

\begin{equation}
        \rho = \rho_0 \rm exp(-\frac{z^2}{2H_{\rm ams}^2}),
\end{equation}

where the initial density at the AGN disk median plane $\rho_0$, in terms of the gravitational Toomre parameter $Q$, is given as 

\begin{equation}\label{equ:rho_Q}
        \rho_0 = \frac{\Omega^2}{\sqrt{2\pi}\pi G Q}.
\end{equation}

We adopt the equation of state as isothermal with a specific heat ratio of $\gamma=1.0$. In our simulations, we choose $Q = 10$ and $\Omega = 10^{-9}\, \rm s^{-1}$, corresponding to the radius $R_{\rm ams} = (GM_{\rm agn}/\Omega^2)^{1/3}$ with respect to the AGN center. Throughout the simulations, any mass accreted to the sink or removed from the grid is not added to the particle, ensuring a constant mass for the sink particle. By selecting different parameters $h$ and $\mu$, we evolve the models until the accretion rate converges to a steady state, and then we measure the mean accretion rate.

It is crucial to differentiate between the concepts of mass feeding rate and accretion rate. The feeding rate indicates the amount of gas that the AGN disk can supply to feed the central AMS, while the accretion rate refers to the mass acquired by the AMS. These rates typically differ because feedback from the AMS disk and the AMS itself can generate outflows, resulting in the accretion rate being usually lower than the feeding rate. In this study, our focus is primarily on the mass feeding rate. However, simulations are conducted for the outer region of the AMS disk, where the impact of stellar feedback is not as significant (see the discussion of feedback in Section \ref{subsect:feedback}). Therefore, disregarding the feedback effect, the mass feeding rate closely corresponds to the mass accretion rate, and in this study, we interchangeably refer to these two terms.

\begin{figure}
    \center 
    \includegraphics[width=0.48\textwidth]{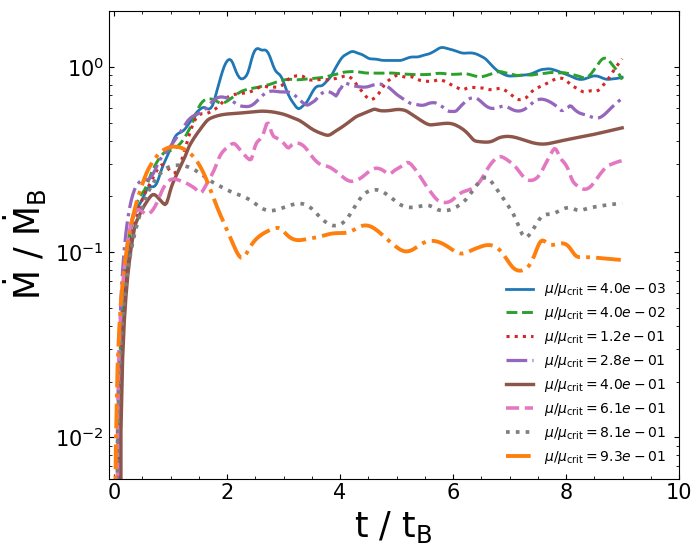}
    \caption{The mass feeding rate on the AMS in units of Bondi rate is plotted as a function of time. Different colors are used to distinguish lines for various values of $\mu/\mu_{\rm crit}$. Results for models with $h = 0.035$ are shown here.}
    \label{fig:rate_time}
\end{figure}

\begin{figure}
    \center 
    \includegraphics[width=0.48\textwidth]{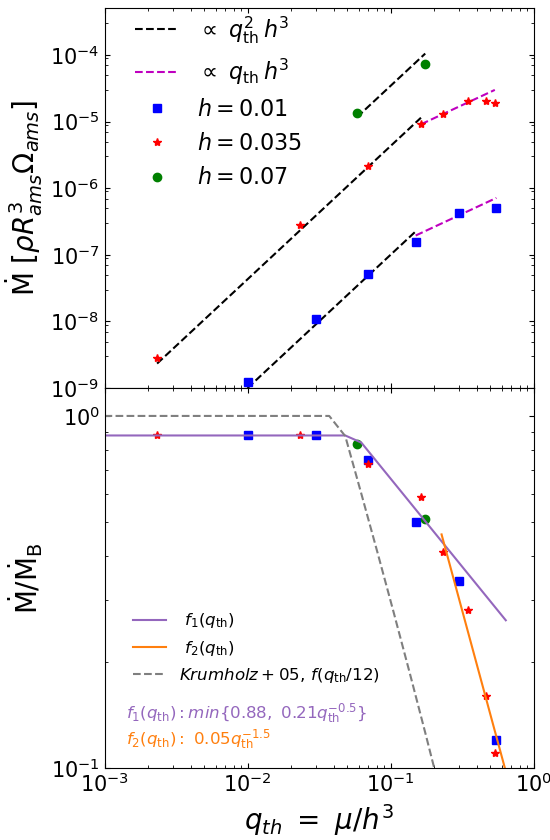}
    \caption{Top panel: The accretion rate, measured in units of $\rho R^3 \Omega$, is plotted against the thermal mass $q_{\rm th}$. Data points are obtained from various AGN disk thickness ratio $h$. The power-law indexes of the dependence are consistent with the results obtained in the planetary accretion models \citep[e.g.,][]{choksi23}.
     Bottom panel:  The dependence of mass feeding rate on the parameter $q_{\rm th}$ is plotted. A solid line represents a fit to the simulation results, while a dashed line shows the formula from \citet{krumholz05} for comparison.}    
    \label{fig:rate}
\end{figure}

\section{Results}
\label{sect:result}

In research concerning the process of material accumulation onto a central point object, the concept of Bondi-like accretion is frequently utilized. The common assumptions of the spherically symmetric Bondi solution prescribe certain values of temperature and density at infinity. However, when considering the impact of angular momentum, the gas tends to form a disk, leading to a potential decrease in the accretion rate compared to the Bondi rate. \citet{proga03} conducted numerical simulations in two dimensions of rotating accretion flow around black holes, demonstrating that as the specific angular momentum increases, the accretion rate decreases. Additionally, \citet{krumholz05} verified that even a small degree of vorticity can significantly alter both the accretion rate and the morphology of the flow lines. They presented an approximate formula describing the accretion rate in relation to the vorticity of the ambient gas. Subsequent research has explored the effects of turbulence by \citet{krumholz06} and viscosity by \citet{narayan11}.

What is the mass feeding rate onto the AMSs? The quantity of gas that gets accreted is influenced by factors such as the vorticity of the surrounding gas, turbulence, viscosity, and radiation feedback from the AMSs. At the formation stage of the AMS in the disk, the effective accretion radius is relatively small. The accretion process resembles Bondi-like accretion with angular momentum. The geometry of the AGN disk and the gravitational pull of the SMBH minimally impact the feeding rate onto the AMS.

\begin{figure*}
    \center 
    \includegraphics[width=0.98\textwidth]{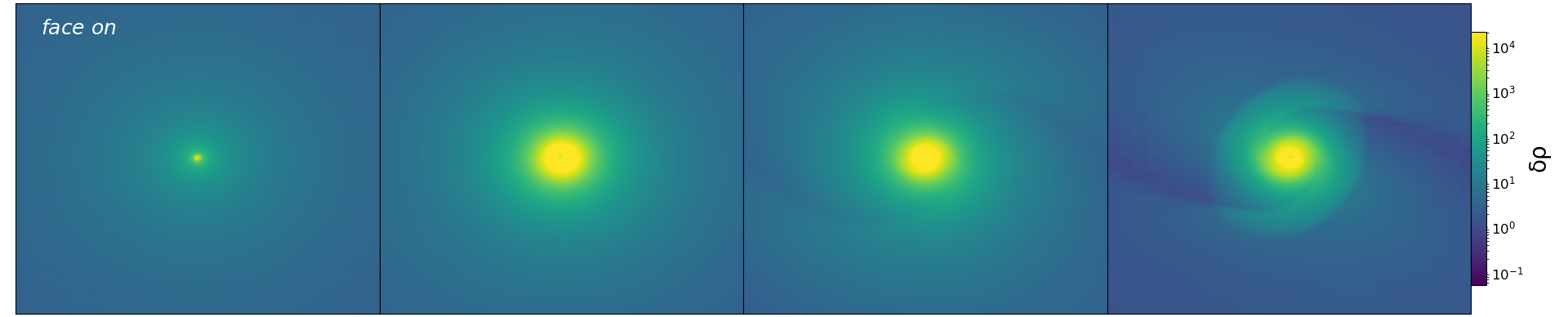}
    \includegraphics[width=0.98\textwidth]{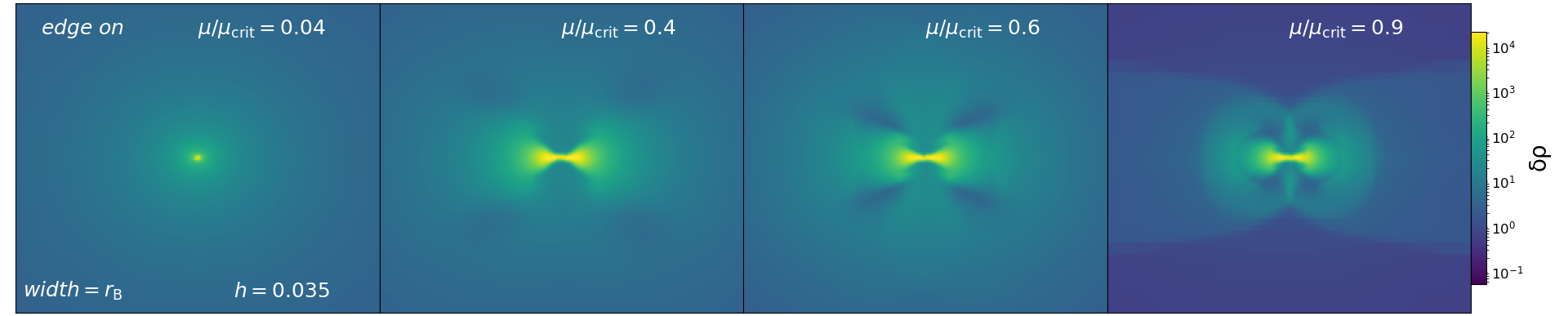}
    \caption{Slices of gas over-density viewed face-on (top panels) and edge on (middle panels), on scale of the the Bondi radius. Here $\delta \rho$ is defined as the ratio of the gas density to the initial midplane density $\rho_0$. Plots are taken from snapshots at the end of the simulation, with the center located on the sink particle. Results for models with $h = 0.035$ are shown here.}
    \label{fig:slice}
\end{figure*}

\begin{figure}
    \center 
    \includegraphics[width=0.48\textwidth]{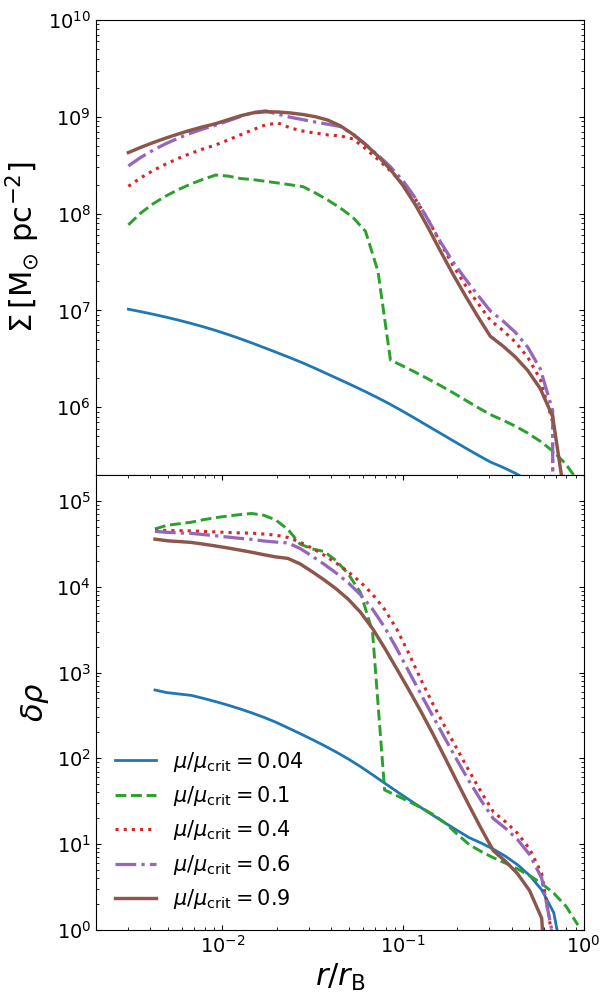}
    \caption{The radial profiles of disk surface density $\Sigma$ and density $\rho$ for various values of $\mu$, after the disk has reached a steady state. Results for models with $h = 0.035$ are shown here.}
    \label{fig:prof}
\end{figure}

In Figure. \ref{fig:rate_time}, the ratio of the accretion rate to the Bondi rate is presented over the evolution time relative to the Bondi timescale. The Bondi timescale, denoted as $t_{\rm B}$, is calculated as $t_{\rm B} = r_{\rm B}/c_{\rm s}$. Results for models with $h = 0.035$ are shown here. 

Our study involves models with varying values of $\mu$, and it could be found that as $\mu$ increases, the ratio of the accretion rate to the Bondi rate decreases. As the mass of the AMS increases, the Bondi radius $r_{\rm B}$ also expands, leading to a rise in the total angular momentum within the gravitational influence radius. During the mass infall process, the accumulated angular momentum towards the center acts as a barrier to gas accretion, causing the accretion rate to deviate from the Bondi rates.

The inflow rate quickly stabilizes, achieving a steady state within $2-3$ Bondi timescales. We have conducted simulations for approximately $\sim 9$ Bondi timescales, and the rate remains stable throughout. As we progress in time, the impact of boundary conditions appears to be less significant in this context. In our current investigation, we constrain the models to cases where $\mu/\mu_{\rm crit}$ is less than one, as values exceeding unity would result in the gravitational tidal influence of the SMBH becoming predominant.

Since the Bondi accretion rate $\dot{M}_{\rm B}$ can be expressed as $\dot{M}_{\rm B} \sim r_{\rm B}^2 \rho c_{\rm s} \sim q_{\rm th}^2 h^3 \rho R_{\rm ams}^3 \Omega$, the top panel of Figure \ref{fig:rate} illustrates the accretion rates in units of $\rho R_{\rm ams}^3 \Omega_{\rm ams}$ as a function of $q_{\rm th}$. Here, $R_{\rm ams}$ and $\Omega_{\rm ams}$ denote the radius and angular velocity of AMS in the AGN disk. We could find that the accretion rates follow the $q_{\rm th}^2$ scaling for $q_{\rm th} < 0.2$, and the $q_{\rm th}$ scaling for $q_{\rm th} > 0.2$. 

The blue squares, red stars, and green circles represent data points obtained from the averaged steady values for models with $h$ values of $0.01$, $0.035$, and $0.07$, respectively. As $q_{\rm th}$ becomes higher than $\sim 0.2$, there is a slight divergence from the $q_{\rm th}^2$ trend in accretion rate. As shown in \citet{choksi23}, when $q_{\rm th}$ becomes larger than $0.3$, the relationship depicted in the top panel approaches $\sim q_{\rm th} h^3$. Our result is consistent with the result obtained in \citet{choksi23}, which works on planetary accretion. At higher $q_{\rm th}$ values, the growth of AMS is limited by SMBH tides, resulting in $\dot{M}$ being less strongly dependent on $q_{\rm th}$.

For a disk that is thermally supported, the scale height is given by $H_{\rm ams} = c_{\rm s}/\Omega$. Consequently, $q_{\rm th}$ can be expressed as $q_{\rm th} = r_{\rm B}/H_{\rm ams} = r_{\rm B}\Omega/c_{\rm s}$. By substituting $r_{\rm B}$, we derive $q_{\rm th} = \mu h^{-3}$. This can also be expressed as $\mu / \mu_{\rm crit} = 3^{1/2}q_{\rm th}$.

For accreting gas with a small amount of angular momentum or vorticity, it has been shown that the reduction in accretion rate could be approximated using the formula $\dot{M}/\dot{M}_{B} = f(q_{\rm th})$, where $f(q_{\rm th}) = \rm min\{1.0, 2(\pi q_{\rm th})^{-1}sinh^{-1}[(2q_{\rm th})^{1/3}]\}$ \citep{krumholz05}. In the bottom panel of Figure. \ref{fig:rate}, we show the ratio of mean accretion rate over Bondi rate after reaching the equilibrium state, as function of $q_{\rm th}$. After applying a least-squares fitting to all $h$ models, the data points could be represented by a broken power law such as

\begin{align}
\label{equ:fomega}
    f_1(q_{\rm th}) &= \rm min\{0.88,\ 0.21 q_{\rm th}^{-0.5\pm0.08}\}  \\
    f_2(q_{\rm th}) &= \rm 0.05 q_{\rm th}^{-1.5\pm0.12}.
\end{align}

In the bottom panel, the derived formulas, are depicted by the solid lines, contrasting with the theoretical formula proposed by \citet{krumholz05} represented by the dashed black line. It is important to note that according to \citet{dittmann21}, the average specific angular momentum within the Bondi radius is $<l> \sim r_{\rm B}^2 \Omega /12$, hence the term $q_{\rm th}$ in the $f(q_{\rm th})$ formula should be adjusted to $q_{\rm th}/12$.

For lower AMS masses where $\mu/\mu_{\rm crit} < 0.5$ or $q_{\rm th} \leq 0.3$, the data aligns with the $f_1(q_{\rm th})$ profile, whereas for $\mu/\mu_{\rm crit} > 0.5$, the data aligns with the $f_2(q_{\rm th})$ profile. The power-law indices for these profiles are presented with a margin of error within $1 \sigma$ significance. When $q_{\rm th}$ drops below approximately $3 \times 10^{-2}$, the rate ratios remain nearly constant at around $0.88$. At $\mu/\mu_{\rm crit} \sim 0.6$, there is an approximate $\sim 15\%$ reduction in the feeding rate relative to the $f_1(q_{\rm th})$ profile. When $\mu/\mu_{\rm crit}$ approaches $0.9$, this rate falls by a factor of $\sim 2$ compared to that described by the $f_1(q_{\rm th})$ profile. This broken at $\mu/\mu_{\rm crit} \sim 0.5$ is anticipated, since as the AMS Bondi radius approaches the disk scale height, the disk geometric influence can limit the mass accretion rate.

For various $h$ values, the difference in the fit formula normalization remains minimal. Thus, the relationship is applicable across all models, regardless of the disk aspect ratio $h$ or the gas sound speed.

In our shearing box simulations, we have incorporated the acceleration from the SMBH in the vertical direction, referring to it as the second term on the RHS of Equations \ref{eq:cons_momentum}. Could this vertical acceleration influence the mass feeding rate? The vertical acceleration is approximately $g_{\rm smbh} = \Omega^2 z$, while the acceleration provided by the AMS is $g_{\rm ams} = GM_{\rm ams}/z^2$. Consequently, the ratio can be expressed as $g_{\rm smbh}/g_{\rm ams} = \mu^{-1}(z/R_{\rm ams})^2$. Assuming $z = y r_{\rm B}$, the ratio becomes $g_{\rm smbh}/g_{\rm ams} = 0.33 y^3 (\mu/\mu_{\rm crit})^2$. Even when both $y$ and $\mu/\mu_{\rm crit}$ approach one, the acceleration due to the SMBH remains only about 30\% of the AMS gravitational attraction, indicating that the gravitational influence of the SMBH is relatively minor.

In Figure \ref{fig:slice}, we present gas density enhancement $\delta \rho$ slices, for models with $h = 0.035$, viewed face-on in the top panels and edge-on in the middle panels, scaled to the Bondi radius. The bottom panels show the absolute density profiles viewed edge-on. The density enhancement $\delta \rho$ is defined as the ratio of the gas density to the initial midplane density $\rho_0$ (equation \ref{equ:rho_Q}). These slices are derived from snapshots taken at the end of the simulation, with the plots centered on the sink particle position. The color bar indicates the ratio of inflow density to the initial density. Moving from left to right panels, we illustrate slices for various $\mu/\mu_{\rm crit}$ models.

For models with a small value of $\mu/\mu_{\rm crit}$, the flow pattern resembles the spherical symmetry of the Bondi flow. Conversely, as $\mu/\mu_{\rm crit}$ approaches one, it results in a more pronounced presence of non-axisymmetric density patterns or density enhancements in the spiral arms. The AMS flow pattern in a hierarchical system exhibits complexity, including streamlines that undergo horseshoe motion, where gas elements execute a U-turn as they approach the central object \citep{ormel13}. In such an environment, the mass feeding rate can deviate from the results obtained by \citet{krumholz05}.

Figure\,\ref{fig:prof} illustrates the radial profiles of the disk surface density $\Sigma$ (upper panel) and density enhancement $\delta \rho$ (lower panel) across various $\mu$ values and models with $h = 0.035$. When $\mu/\mu_{\rm crit} = 0.04$, the gas inflow exhibits nearly spherical symmetry, and the density profile follows a $\rho \sim r^{-1}$ trend. With increasing $q_{\rm th}$, gas accumulates at the center, leading to the formation of an accretion disk. For $\mu/\mu_{\rm crit} = 0.1$, the accretion disk size is approximately $\sim 0.08\, r_{\rm B}$. Increased gas accumulation leads to significant rises in both density and surface density. The density profile becomes flat within $\sim 0.04\, r_{\rm B}$. As angular momentum or $q_{\rm th}$ rises, the disk surface density within $\sim 0.1\, r_{\rm B}$ increases.

In our models, when $\mu/\mu_{\rm crit}$ exceeds $0.1$, the disk size is approximately $\sim 0.1\, r_{\rm B}$. The aspect ratio of disk thickness ranges approximately from $0.2$ to $1$, with the disk becoming slightly thinner as the value of $\mu/\mu_{\rm crit}$ increases. The density ratio $\delta \rho$ near the center can reach as high as $\sim \rm few \times 10^4$, whereas it is only around $\sim 10^3$ in \citet{krumholz05}, such that our models exhibit a more concentrated gas distribution, resulting in a slightly higher accretion rate towards the AMS.

%


\section{Discussions}
\label{sect:discuss}
\subsection{Feedback during AMS growth}
\label{subsect:feedback}

During the growth of the AMS, various feedback effects can occur, including both radiative and mechanical influences. Initially, we focus on the radiative feedback to assess its significance. Radiation can release from the rapid accretion process and the nuclear burning within the AMS itself. In the case of the accretion process, the accretion rate initially resembles the Bondi rate, but it is adjusted by factors such as gas angular momentum and turbulence.

During the early stages of AMS growth, as presented in the Section \ref{sect:result}, the adjusted mass feeding rate can be approximated as $\dot{M} \sim f(q_{\rm th}) \dot{M}_{\rm B}$. The expression for the accretion luminosity, if assuming optically thin for the accretion flow, is roughly given by $L_{\rm acc} \sim GM_{\rm ams}\dot{M}/r_{\rm ams}$, where $r_{\rm ams}$ denotes the radius of the AMS. To obtain the AMS radius, by assuming a zero-age main sequence star with solar metallicity, one can employ the approach described in \citet{bond84}, resulting in $r_{\rm ams}/r_{\odot} \sim 0.56 (M_{\rm ams}/M_\odot)^{0.51}$, where $r_{\odot}$ is the solar radius. Hence, the accretion luminosity can be computed as:

\begin{equation}
 L_{\rm acc} \sim 1.4\times 10^{46}\,f(q_{\rm th})\, \frac{m_{\rm bondi}}{10^9}\, \left( \frac{M_{\rm ams}}{10^3\, M_\odot} \right)^{1.49} \,\rm erg\,s^{-1},
\end{equation}

where $m_{\rm bondi}$ represents the ratio of AMS Bondi accretion rate to the Eddington rate $\dot{M}_{\rm edd}$, as discussed in \citet{wang21}, with $m_{\rm bondi}$ possibly reaching as high as $\sim 10^9$. It is important to highlight that, when it comes to the process of accretion onto a star, the Eddington luminosity and accretion rate, $L_{\rm Edd}\sim \dot M_{\rm Edd, ams} v_{\rm ff}^2$, are linked through $v_{\rm ff}^2\sim G M_{\rm ams}/r_{\rm ams}$, which implies that $\dot{M}_{\rm Edd,ams} \sim 10^{-3} r_{\rm ams}/r_\odot\ \rm M_\odot\,yr^{-1}$. In our study, the accretion rates, for certain modeled parameters, still remain comparable to or higher than $\dot{M}_{\rm Edd,ams}$. Therefore, the accretion luminosity directed towards an AMS could be similar to or lower than the Eddington luminosity because of photon trapping. Here, we continue to employ $m_{\rm bondi} = \dot{M}_{\rm B}/\dot{M}_{\rm Edd}$ for simplicity.

It is also important to note that if the accretion time scale is significantly shorter than the thermal adjustment timescale, the radius can expand significantly leading to a decrease in the effective temperature \citep{woods21,herrington23,woods24}. This phenomenon occurs because the gas flow could not efficiently release radiation, causing energy accumulation and subsequent bloating of the surface \citep{schleicher13}. Therefore, when the short accretion timescale is taken into account, the accretion luminosity might be lower than expected.

The radiative luminosity from the AMS due to nuclear burning can be approximated by the Eddington luminosity. For stellar masses below $10\, M_\odot$, the luminosity may be lower than the Eddington luminosity, but as the AMS increases in mass, radiation pressure in the stellar interior becomes dominant, resulting in a final luminosity close to the Eddington limit \citep{woods21,woods24}. In comparison to the accretion luminosity, which surpasses the Eddington limit by a significant margin, most of the total radiation is primarily contributed by accretion.

Could radiation from the accretion flow and the AMS nuclear burning have any impact on the accretion flow? To show this, we compare the momentum in radiation with the momentum of the accretion flow. These two terms are given as 

\begin{align}
L_{\rm acc}/c &\sim 4.8\times 10^{35}\,f(q_{\rm th})\, \frac{m_{\rm bondi}}{10^9}\, \left( \frac{M_{\rm ams}}{10^3\, M_\odot} \right)^{1.49} \rm erg\,cm^{-1}, 
\\
\dot{M}v &\sim \dot{M} \sqrt{GM_{\rm ams}/R_{\rm ams}} \\
         &\sim 4.6\times 10^{37}\,f(q_{\rm th})\, \frac{m_{\rm bondi}}{10^9}\, \left( \frac{M_{\rm ams}}{10^3\, M_\odot} \right)^{1.245} \rm erg\,cm^{-1}.
\end{align}

It becomes apparent that the momentum of the accretion flow consistently dominates the radiative feedback, preventing the radiation from pushing aside the inflow medium \citep[for an alternative aspect of the feedback effect, see][]{chen24b}. 

In addition, there is the impact of radiation trapping. When radiation becomes trapped within the accretion flow, the accreting material loses its ability to radiate energy within $r_{\rm trap}$ \citep{begelman78}. Consequently, it releases energy at a rate $L_{\rm acc} \sim GM_{\rm ams}\dot{M}/r_{\rm trap}$. The trapping radius can be described as

\begin{align}
  r_{\rm trap} &= \frac{\dot{M} \sigma_{\rm T}}{4\pi m_{\rm p}c} \\
          &\sim 2.2\times 10^{12}\,r_\odot\,f(q_{\rm th})\, \frac{m_{\rm bondi}}{10^9},
\end{align}

where $\sigma_{\rm T}$, $ m_{\rm p}$ are the Thomson cross section and the proton mass, respectively. The trapping radius is several orders of magnitude larger than the AMS radius, leading to a substantial reduction in the radiative feedback.

What about the mechanical feedback? When comparing the momentum in radiation and accretion, it is unlikely that outflow could be solely triggered by radiation. According to numerical simulations, outflow from super-critical accretion flow is mostly driven by magnetic force. However, there is skepticism regarding whether the outflow rate could surpass the inflow rate, leading to the possibility that the momentum in the outflow might overwhelm that of accretion.

Another potential source of mechanical feedback is the mass loss from the AMS surface. The mass loss rates of stars can be estimated using the empirically calibrated formula by \citet{lamers93} as:

\begin{equation}
  {\rm log}(\dot{M}_{\rm loss}) = 1.738 {\rm log}(L/L_{\odot}) - 1.352 {\rm log}(T_{\rm eff}) - 9.547.
\end{equation}

Assuming the effective temperature of the AMS as $T_{\rm eff} = 10^5\, K$ and the luminosity $L$ as the Eddington luminosity, the mass-loss rate can be estimated as

\begin{equation}
  \dot{M}_{\rm loss} \sim 6.3\times10^{-4}\, \left( \frac{M_{\rm ams}}{10^3\, M_\odot} \right)^{1.738}\,\rm M_\odot\,yr^{-1}.
\end{equation}

However, this mass-loss rate is significantly lower than the accretion rate shown in Fig. \ref{fig:rate_time}. Therefore, it can be concluded that during the early stage of AMS growth, neither radiative nor mechanical feedback can impede the accretion process.

Previous studies have shown that star mass loss can terminate AMS growth when the loss rate balances the inflow rate \citep{cantiello21,ali-dib23}. The differences between our findings and those of others can be attributed to the birthplace of the AMS or the environment in which it is located. In the study by \citet{cantiello21}, the density of the AGN disk at its mid-plane was estimated to be around $\rho \sim 10^{-16}\,\rm g\,cm^{-3}$, placing the AMS at a distance of about $R_{\rm ams} \sim \rm pc$. In our investigation, we placed the AMS at a radius of $R_{\rm ams} \sim 10^4 R_{\rm g} \sim 0.05\,\rm pc$. Variances in the AGN disk density at different radii could result in varying AMS Bondi rates. Within the AGN disk, there is a crucial radius, where the AMS feeding rate is consistently greater than the mass-loss rate.  Through a series of algebraic manipulations of equations \ref{equ:bondi} and \ref{equ:rho_Q}, this critical radius can be approximated using the formula:

\begin{equation}
 R_{\rm ams, crit} \sim 4.5\times 10^5\ R_{\rm g}\,\left( \frac{\dot{M}_{\rm loss}}{10^{-3}\,\rm M_\odot\,yr^{-1}}\right)^{-2/3}\, \left( \frac{M_{\rm ams}}{10^3\,\rm M_\odot} \right)^{4/3},
\end{equation}
where we have assumed $h = 0.03, \alpha_{\rm B} = 10, f(q_{\rm th}) = 0.1, Q = 10$ and $M_{\rm agn} = 10^8\,\rm M_\odot$.

Finally, we can anticipate an AMS growing without being influenced by feedback within a radius of $R_{\rm ams, crit}$. For AMSs located outside this radius, their growth could be terminated by the AMS wind. However, since the structure of the AMS remains uncertain, there is ongoing discussion regarding the precise rate of mass loss through wind.

All the aforementioned considerations are made under the assumption that the flow resembles a spherical symmetry environment. As the AMS evolves over time, the low-angular momentum gas in the polar region of the AMS disk rapidly moves inward, leading to the cleaning of gas. Consequently, the mass inflow rate along the polar region decreases. Eventually, radiation or outflow may escape through the polar funnel region, but these feedback mechanisms are unlikely to significantly impact the inflow in the mid-plane region.

\subsection{the tidal effect}
\label{subsect:tidal}
For AMSs situated within AGN disks, the gas in proximity to the AMSs may experience tidal forces. The interaction between the AMS and the gaseous AGN disk can result in an exchange of angular momentum, potentially leading to a drop in surface density near the corotation radius, forming a gap, and causing AMS orbital migration \citep{goldreich80}.

Two conditions are necessary for gap formation: first, the Hill sphere (or Roche radius) of the AMS must be comparable to the thickness of the gas disk. Second, from a viscous perspective, the tidal torques must be able to clear the gas from the gap region faster than viscosity can refill it \citep{goldreich80,armitage07,kley12,li23}.

Setting the opening timescale for the gap $t_{\rm open}$ equals the closing timescale $t_{\rm close}$, the equations of which have been listed in Section \ref{sect:expect}. We can obtain a critical mass ratio as

\begin{equation} \label{equ:gap_mu}
    \mu^2 \sim \frac{\alpha c_{\rm s}^2}{R_{\rm agn}^2 \Omega^2 m^2} \sim \alpha h^2/m^2,
\end{equation}

where we have assumed $\nu = \alpha c_{\rm s}^2/\Omega$. For typical parameters $\alpha = 0.1$, $h = 0.03$ and $m \sim 1/h$, the condition for gap opening is satisfied for $\mu = 2.8\times 10^{-4}$, corresponding to an AMS mass of $M_{\rm ams} \sim 2.8\times10^{4}\,M_\odot$. This value of $\mu$ already exceeds the critical value $\mu_{\rm crit} = 1.6\times 10^{-5}$. Consequently, the AMS can achieve $\mu_{\rm crit}$ before the gap is formed due to tidal torque. To create a gap of width $\Delta R/R_{\rm agn} \sim 0.01$ with a mass of $\mu = \mu_{\rm crit}$, the estimated timescale is approximately $t_{\rm open} \sim 1.2\times 10^4\,\rm yr\ \Omega/10^{-9}\,\rm s^{-1}$.

The stellar lifetime of a main-sequence star is directly proportional to its stellar mass divided by its luminosity \citep{prialnik09}. According to the estimates in \citet{schaerer02}, a star with a mass of approximately $\sim 10^3\,M_\odot$ has a lifespan of around $2$ million years ($t_{\rm star} \sim 2\,\rm Myr$), which is notably longer to the timescale required to open a gap, assuming the AMS is massive. Consequently, the AMS would explode as supernovae after the gap forms. For an accreting star, the stellar lifetime could also be extended due to the recycling of hydrogen-rich disk gas, but it depends on the accretion rate and outflows \citep{chen24}.

\section{Concluding Remarks}
\label{sect:conclusion}

In this research, we performed $3D$ numerical simulations to calculate the mass feeding rates onto stars which are embeded within AGN disks.

\begin{itemize}
    \item The shearing of the AGN disk with angular momentum can notably decrease the mass supply rates towards the AMS. We have presented an approximate formula that explains the mass feeding rate according to the angular momentum of the surrounding gas.

    \item When examining the scalability of our AMS accretion models in relation to planet formation, we have observed that the accretion processes involved in planet formation exhibit similarities to AMS accretion. We have verified that the maximum accretion rates towards an AMS adhere to the same relationships as those involved in early planet accretion.

\end{itemize}

The equation \ref{equ:fomega} remains valuable for theoretical understanding the initial growth of the AMS. The timescale for gap-opening resulting from tidal torque is inversely proportional to the square of the AMS mass, $t_{\rm open} \sim \mu^{-2}$. Consequently, if the AMS experiences rapid growth, the gap-opening timescale $t_{\rm open}$ will decrease significantly, facilitating the formation of a gap at an earlier stage. By considering the interaction between the AGN disk mass feeding (constrained by viscosity $\nu$) and the gap-opening process (governed by the mass $\mu$), it prompts the question of how long the AMS can maintain its growth phase. In this study, we focus on analyzing AMS growth with small $\mu$, while the future evolution of AMS will be explored in a future research.

Inspired by the observed top-heavy present-day mass function close to the galactic SMBH \citep[e.g,][]{nayakshin05,lu13,hosek19}, it is suggested that intermediate-mass stars might have formed in the SMBH accretion disk during a earlier accretion episode \citep{hocuk11,mapelli12,davies20}. Early research by \citet{nayakshin07} demonstrated that the mass function peaks at roughly $50\, \rm M_\odot$ at the high-mass end. Recent work by \citet{hopkins24} indicates that stars formed in the SMBH disk experience a rapid mass increase, reaching a maximum on a timescale as short as a few years, after which the rate of mass growth significantly slows. \citet{hopkins24} noted that non-thermal factors (such as turbulence, magnetism, star-gas interactions) might play a significant role in limiting the mass growth. Furthermore, in these studies, stars are simulated with feedback from main-sequence stars, whereas for stars with rapid accretion, their internal structure or radiative efficiency needs more thorough examination. Future studies incorporating more detailed models of mass growth suppression under various physical conditions are undoubtedly required.

\section*{Acknowledgements}

We are grateful to the anonymous reviewer for their valuable comments, which have helped improve the quality of our manuscript.
All the analysis has been conducted using yt \citep{turk11}, http://yt-project.org/. 
Y.L. acknowledges the support from NSFC grant No. 12273031. 
J.M.W. acknowledges financial support from the National Key R\&D Program of China (2021YFA1600404 and 2023YFA1607904), the National Natural Science Foundation of China (NSFC; 11833008, 11991050, 11991054, and 12333003).
The simulations were carried out at the National Supercomputer Center in Tianjin, performed on TianHe-1A. Part of the numerical computations were also conducted on the Yunnan University Astronomy Supercomputer.

\section*{Data Availability}

The data underlying this article will be shared on reasonable request to the corresponding author.


\bibliography{ms}{} 
\bibliographystyle{mnras}







\label{lastpage}
\end{document}